\documentclass[conference]{IEEEtran}
\IEEEoverridecommandlockouts
\usepackage{cite}
\usepackage{amsmath,amssymb,amsfonts}
\usepackage{algorithmic}
\usepackage{graphicx}
\usepackage{textcomp}
\usepackage{xcolor}
\usepackage{multirow}
\usepackage{siunitx}

\def\BibTeX{{\rm B\kern-.05em{\sc i\kern-.025em b}\kern-.08em
    T\kern-.1667em\lower.7ex\hbox{E}\kern-.125emX}}
\begin{document}

\title{Low-Rate Wrist SpO$_2$ Estimation under Micro-Perturbations Using Motion-Aware Beat Selection and Perfusion-Guided Calibration\\

}
\author{
\IEEEauthorblockN{
Zequan Liang$^{1}$,
Ning Miao$^{2}$,
Wei Shao$^{1}$,
Mahdi Pirayesh Shirazi Nejad$^{2}$,\\
Ehsan Kourkchi$^{2}$,
Hossein Sayadi$^{3}$,
Setareh Rafatirad$^{1}$,
Houman Homayoun$^{2}$
} \\
\IEEEauthorblockA{
$^{1}$Department of Computer Science, University of California, Davis, Davis, CA, U.S.A.\\  
$^{2}$Department of Electrical and Computer Engineering, University of California, Davis, Davis, CA, U.S.A.\\ 
$^{3}$Department of Computer Engineering and Computer Science, California State University, Long Beach, Long Beach, CA, U.S.A.\\ 
Email: \{zqliang, nmiao, wayshao, pirayesh, ekay\}@ucdavis.edu, hossein.sayadi@csulb.edu, \{srafatirad, hhomayoun\}@ucdavis.edu}} 

\maketitle

\begin{abstract}
Continuous oxygen saturation (SpO$_2$) monitoring from photoplethysmography (PPG) is important for wearable health sensing, but wrist-based SpO$_2$ estimation remains challenging due to subtle wrist micro-perturbations and inter-subject differences in local perfusion status. These factors can destabilize the red-to-infrared ratio-of-ratios (R) and reduce the reliability of conventional fixed R--SpO$_2$ mapping. This paper proposes a lightweight low-rate wrist SpO$_2$ estimation framework that integrates motion-aware beat selection and perfusion-guided calibration. The proposed method extracts beat-level alternating-current/direct-current (AC/DC) components from dual-wavelength PPG signals, computes beat-level R values, and uses accelerometer-derived motion scores to weight beats within each sliding window. A subject-specific perfusion reference is further used to guide calibration across different perfusion conditions. Experiments on a private wearable dataset show that the proposed method achieves the best 25 Hz performance, with an MAE of 2.305 ± 1.113\% and an RMSE of 3.117 ± 1.743\%, while maintaining performance comparable to the 100 Hz sampling rate and reducing PPG sensor power consumption for energy-efficient wearable implementation. These results demonstrate the effectiveness of the proposed framework for low-rate wrist SpO$_2$ estimation under micro-perturbations.
\end{abstract}


\section{Introduction}

\begin{figure}[t]
    \centering
    \includegraphics[width=\linewidth]{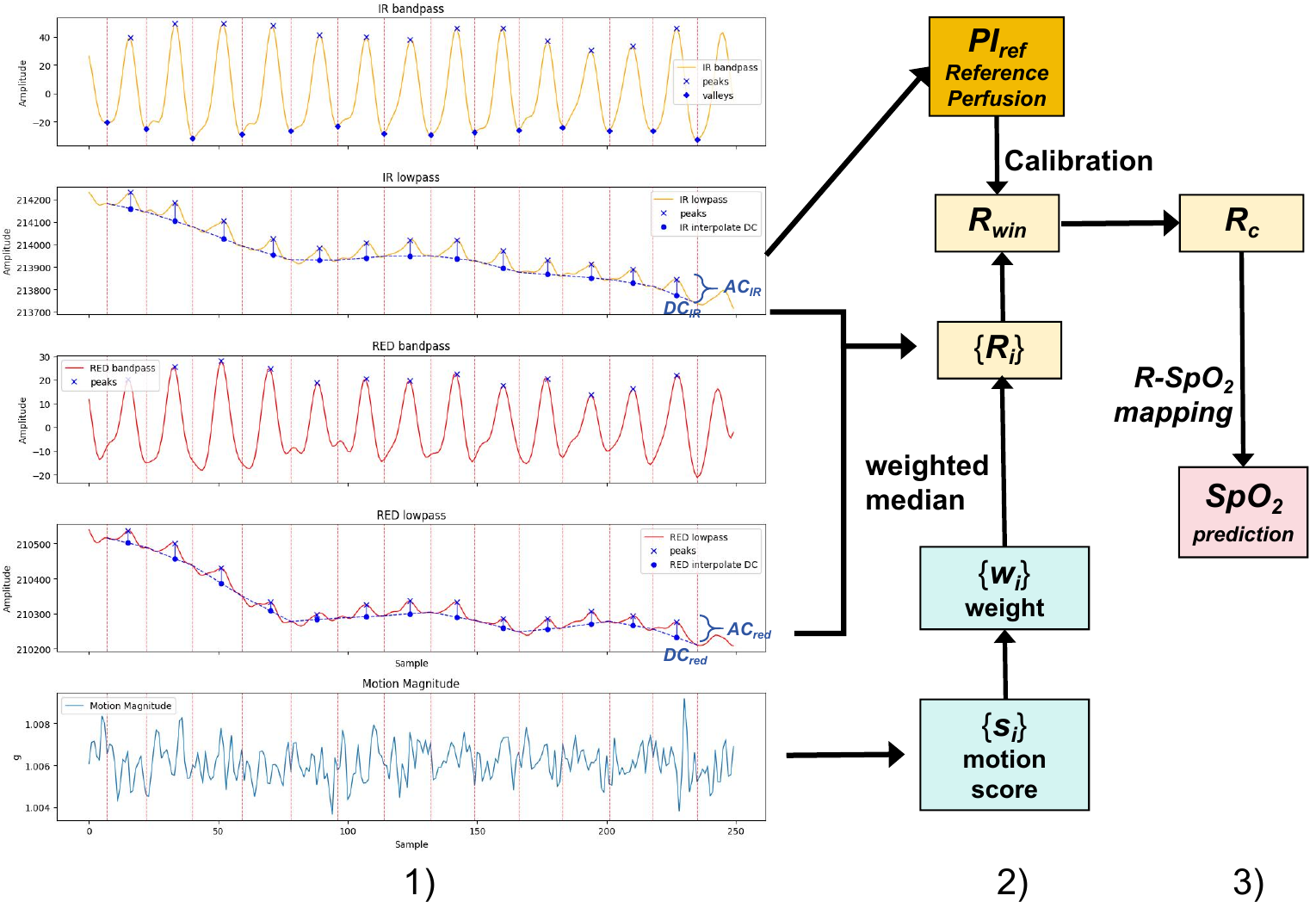}
    \caption{Overview of SpO$_2$ estimation framework. 1) Beat-level AC/DC extraction and Ratio-of-Ratios Computation, 2) Motion-Aware Beat Selection, 3) Perfusion-Guided R–SpO2 Calibration}
    \label{fig:spo2_framework_flowchart}
\end{figure}
Oxygen saturation (SpO$_2$) is an important vital sign for assessing respiratory and cardiovascular status \cite{evans2001vital}. However, wrist-based SpO$_2$ estimation remains more challenging than conventional finger-based pulse oximetry. Compared with the finger, the wrist is more susceptible to small movements, contact pressure changes, and weaker pulsatile perfusion \cite{phillips2021wristo2,DBLP:conf/ndss/ShaoLZFMKRHF26}.

In practical wrist-based SpO$_2$ sensing, small wrist movements and subtle contact changes can perturb the morphology of PPG beats \cite{shao2025self}. These micro-perturbations may alter the red and infrared pulsatile components, leading to instability in the red-to-infrared ratio-of-ratios (R). Since conventional SpO$_2$ estimation relies on a fixed calibration curve between R and SpO$_2$, beat-level instability in R can directly produce inconsistent SpO$_2$ estimates.

Another source of variability comes from differences in local perfusion conditions across participants~\cite{scardulla2020study}. The perfusion index baseline may vary due to individual physiological differences and skin--sensor contact. As a result, a conventional fixed R--SpO$_2$ mapping may introduce calibration bias under different perfusion states.

Motivated by these observations, this work treats wrist-based SpO$_2$ estimation as a micro-perturbation and perfusion-sensitive calibration problem. Instead of relying only on a fixed R--SpO$_2$ mapping, we introduce a lightweight processing strategy that combines motion-aware beat selection and perfusion-guided calibration. Accelerometer information is used to identify more reliable beats under lower motion, while the perfusion index is used as a calibration-state indicator to adapt the R-SpO$_2$ relationship across various subjects. This design aims to improve the robustness of low-rate wrist SpO$_2$ estimation without requiring a high-complexity denoising model.

As shown in Fig.~\ref{fig:spo2_framework_flowchart}, the main contributions of this work are summarized as follows:

\begin{itemize}
    \item We propose a low-rate dual-wavelength PPG processing algorithm for wrist-based SpO$_2$ estimation at 25 Hz. The method performs beat-by-beat extraction and computes multiple red-to-infrared ratio-of-ratios values within a rolling window for subsequent R-SpO$_2$ calibration.

    \item We introduce a motion-aware beat selection strategy to improve robustness under micro-perturbations. Accelerometer-derived motion information is used to assign reliability weights to beats within each window, so that beats collected under lower-motion conditions contribute more strongly to the R value estimation.
    
    \item We propose a perfusion-guided R--SpO$_2$ calibration method that uses the perfusion index as a calibration-state indicator. This allows the R--SpO$_2$ relationship to adapt to subject-specific variability in local perfusion conditions.

\end{itemize}

Overall, this work presents a lightweight and firmware-compatible framework for low-rate wrist SpO$_2$ estimation under micro-perturbations and varying perfusion conditions.

\section{Methodology}

\subsection{Overview}

The proposed framework estimates SpO$_2$ from low-rate wrist-worn dual-wavelength PPG signals. 
The input signals include red PPG, infrared (IR) PPG, and tri-axial accelerometer measurements. 
As shown in Fig.~\ref{fig:spo2_framework_flowchart}, 
it consists of three main steps: 
1) beat-level alternating-current/direct-current (AC/DC) extraction
 to compute perfusion indices (PI) and the red-to-infrared ratio-of-ratios (R) \cite{guo2015reflective}, 
2) motion-aware beat selection using accelerometer-derived reliability weights, and 
3) perfusion-guided R--SpO$_2$ calibration.

\subsection{Beat-Level AC/DC Extraction and Ratio-of-Ratios Computation}

For each 10-s sliding window, the red and infrared (IR) PPG signals are processed independently. 
Each PPG channel is first low-pass filtered with a 5 Hz cutoff to preserve the pulse morphology while suppressing high-frequency noise. 
In parallel, a 1--3 Hz band-pass filtered signal is generated to emphasize the cardiac pulsatile component and facilitate robust beat detection.

Peak and valley detection is performed on the band-pass filtered PPG signals. Since the IR PPG signal provides more stable beat boundary information in our recordings, valleys detected from the IR channel are used as the common beat boundaries for both wavelengths. The systolic peak is then identified separately from the red and infrared band-pass signals to obtain the beat-level pulsatile maxima. The detected valley and channel-specific peak locations are mapped back to the 0--5~Hz low-pass PPG waveforms for AC/DC extraction.

Given two adjacent valleys, a linear baseline is constructed between their amplitudes in the low-pass waveform. 
The interpolated baseline value at the peak location is defined as the beat-level DC component. 
The beat-level AC component is then computed as the amplitude difference between the peak value and this interpolated baseline. 
This procedure is applied independently to the red and IR channels.

The beat-level red-to-infrared ratio-of-ratios is then computed from the PI of the two wavelengths as
\begin{equation}
R = \frac{PI_{red}}{PI_{IR}}
= \frac{AC_{red}/DC_{red}}{AC_{IR}/DC_{IR}}.
\end{equation}

Since the proposed framework uses a 10-s sliding window with a 1-s step size, multiple valid beats can be extracted within each window. 
The resulting beat-level R values are then passed to the motion-aware beat selection and perfusion-guided calibration stages.

\subsection{Motion-Aware Beat Selection}

To reduce the influence of micro-perturbations, beat-level ratio-of-ratios values are aggregated using accelerometer-derived reliability weights. 
For the $i$-th beat interval $\mathcal{B}_i$, the motion magnitude is computed from the tri-axial accelerometer signals, and its standard deviation is used as the beat-level motion score:
\begin{equation}
s_i = \mathrm{std}\left(\left\{\sqrt{a_x[n]^2 + a_y[n]^2 + a_z[n]^2} \mid n \in \mathcal{B}_i\right\}\right),
\end{equation}
where $a_x[n]$, $a_y[n]$, and $a_z[n]$ denote the three accelerometer axes. 

Within each 10-s sliding window, multiple valid beats can be detected, resulting in a set of beat-level motion scores $\{s_i\}$ and ratio-of-ratios values $\{R_i\}$. 
The distribution of the motion scores in the current window is summarized using two percentile-based thresholds, denoted as $s_{\mathrm{low}}$ and $s_{\mathrm{high}}$. 
In this study, these thresholds are set as the median and the 90th percentile of the beat-level motion scores. 
Each beat is then assigned a nonlinear reliability weight according to
\begin{equation}
u_i = \mathrm{clip}\left(
\frac{s_i - s_{\mathrm{low}}}{s_{\mathrm{high}} - s_{\mathrm{low}}},
0, 1
\right),
\end{equation}
\begin{equation}
w_i = (1-u_i)^{\gamma},
\end{equation}
where $\gamma=3$ in this study is used to emphasize low-motion beats while suppressing motion-contaminated beats.

Finally, the window-level motion-aware ratio-of-ratios value $R_{\mathrm{win}}$ is obtained by applying a weighted median to the valid beat-level $R_i$ values \cite{brownrigg1984weighted}:
\begin{equation}
R_{\mathrm{win}} = R_{(k)}, \quad
k = \min \left\{ j : \sum_{i=1}^{j} w_{(i)} \geq \frac{1}{2}\sum_{i=1}^{N} w_{(i)} \right\},
\end{equation}
where $R_{(i)}$ denotes the beat-level ratio-of-ratios values sorted in ascending order, and $w_{(i)}$ denotes the corresponding reliability weights after sorting.

\subsection{Perfusion-Index-Guided R--SpO$_2$ Calibration}

To account for subject-specific differences in optical coupling and peripheral perfusion, a reference perfusion index, $PI_{\mathrm{ref}}$, is estimated for each session. 
Specifically, we searched the first 30-s segment to identify the window with the lowest average accelerometer-derived motion score, which is treated as the most reliable reference window. 
The reference perfusion index is then computed as the weighted median of the corresponding infrared perfusion index values, $\{PI_{\mathrm{IR}}\}$, using the same motion-aware weighting strategy as that used for estimating $R_{\mathrm{win}}$. 

The window-level ratio-of-ratios value $R_{\mathrm{win}}$ is then corrected using the subject-specific reference PI as
\begin{equation}
R_c
=
R_{\mathrm{win}} + b_0 + b_1 \cdot PI_{ref},
\end{equation}
where $b_0$ and $b_1$ are calibration coefficients learned from the training data. 
The corrected ratio $R_c$ is finally mapped to SpO$_2$ through a second-order calibration function:
\begin{equation}
\widehat{SpO}_2
=
a_2 R_c^2 + a_1 R_c + a_0,
\end{equation}
where $a_0$, $a_1$, and $a_2$ are polynomial calibration coefficients learned from the training data. 
The resulting $\widehat{SpO}_2$ is the final estimated oxygen saturation for the current sliding window.

During training, the ground-truth SpO$_2$ labels were used only to learn the perfusion-guided R--SpO$_2$ calibration parameters. During testing, no SpO$_2$ labels were used, $PI_{\mathrm{ref}}$ was estimated directly from the test PPG recording and applied to the learned calibration model.

\section{Experiment Results}

\subsection{Dataset}

We collected a private wearable dataset from nine participants with diverse skin tones. 
The dataset includes dual-channel 100 Hz PPG signals from the We-Be band, along with corresponding SpO2 reference values from Masimo Rad-G. The reference SpO$_2$ values ranged from 60\% to 100\%.
In total, we collected 27 sessions from 9 individuals, with 3 sessions per person.  Each participant performed 3 breath-holding trials to induce transient SpO$_2$ fluctuations during each session. For low-rate evaluation, the 100 Hz PPG signals were downsampled to 25 Hz.
Further details of the data collection setup and device configuration follow our previous wearable SpO$_2$ data collection protocol~\cite{liang2025rapid}.

\subsection{Comparison of Window-Level R Computation Methods}

We evaluated different strategies for computing the window-level ratio-of-ratios value $R_{\mathrm{win}}$. 
All results were obtained under a strict leave-one-subject-out (LOSO) evaluation. 
For each test subject, we computed the mean absolute error (MAE), root mean squared error (RMSE), Pearson correlation coefficient (PCC), RMSE for SpO$_2 \geq 85\%$, and RMSE for SpO$_2 < 85\%$ between the predicted and ground-truth SpO$_2$ values. The reported results were averaged across all test subjects.

As shown in Table~\ref{tab:spo2_r_value_method_ablation}, we compared the proposed motion-aware weighted method with several baseline strategies for computing $R_{\mathrm{win}}$. 
All methods used the same perfusion-guided calibration module and differed only in how $R_{\mathrm{win}}$ was obtained.
The mean baseline directly averages all valid beat-level R values within the window, while the median baseline uses the unweighted median. 
In addition, Recursive Least Square \cite{mugdha2015study} (RLS) and Normalized Least Mean Square (NLMS) \cite{rupp2002behavior} are evaluated as adaptive filtering methods that use accelerometer information to reduce motion contamination in the PPG signals before ratio-of-ratios extraction.

Under the same perfusion-guided calibration module, the proposed motion-aware weighted aggregation achieves the best overall performance among all compared approaches, with an MAE of 2.305 ± 1.113\%, an RMSE of 3.117 ± 1.743\%.
Compared with the mean, median, RLS, and NLMS baselines, the proposed method achieves better performance by using accelerometer-derived reliability weights to select more reliable beats for window-level $R_{\mathrm{win}}$ estimation.

\begin{table}[t]
\caption{SpO$_2$ Estimation Performance under Different $R_{win}$ Computation Methods}
\setlength{\tabcolsep}{5.0pt}
\begin{center}
\begin{tabular}{|c|c|c|c|c|c|}
\hline
\textbf{Method} 
& \textbf{MAE} 
& \textbf{RMSE} 
& \textbf{PCC}
& \textbf{RMSE $\geq$85\%}
& \textbf{RMSE $<$85\%} \\
\hline

Mean 
& 2.936 & 4.161 & 0.590 & 4.020 & 7.828 \\
\hline

Median 
& 2.370 & 3.228 & 0.663 & 3.108 & 6.960 \\
\hline

RLS 
& 2.403 & 3.255 & 0.663 & 3.139 & 6.988 \\
\hline

NLMS 
& 2.397 & 3.248 & 0.674 & 3.123 & 7.035 \\
\hline

Ours 
& \textbf{2.305} & \textbf{3.117} & \textbf{0.685} & \textbf{3.007} & \textbf{6.804} \\
\hline

\end{tabular}
\label{tab:spo2_r_value_method_ablation}
\end{center}
\end{table}

\subsection{Ablation Study and Low-Rate Evaluation}

We further conducted an ablation study to evaluate the contribution of motion-aware beat selection and perfusion-guided calibration under different sampling rates. When motion-aware beat selection (Motion.) was removed, multiple beat-level R values within each window were aggregated using the unweighted median to obtain $R_{\mathrm{win}}$. When perfusion-guided calibration (Perfusion.) was removed, the estimated $R_{\mathrm{win}}$ was directly mapped to SpO$_2$ using the conventional polynomial R--SpO$_2$ calibration function. 

Table~\ref{tab:spo2_sampling_input_ablation} shows that both proposed components are effective. 
Motion-aware beat selection improves the robustness of window-level $R_{\mathrm{win}}$ estimation by down-weighting motion-contaminated beats, while perfusion-guided calibration improves the subject-specific R--SpO$_2$ mapping, particularly under low-saturation conditions. 
Combining both components leads to the best overall performance across most metrics at both sampling rates.

In addition, the proposed method achieves comparable performance at 25 Hz and 100 Hz, indicating that the framework can preserve SpO$_2$ estimation accuracy under low-rate wrist-worn PPG acquisition. In our previous work~\cite{liang2025rapid}, we tested the power consumption of the PPG sensor on the We-Be band. After subtracting the static power drawn by non-PPG components, the PPG sensor consumed approximately 9~mW at 25~Hz, 40\% lower than the approximately 15~mW consumed at 100~Hz. This lower power consumption can extend battery life and support longer-term continuous monitoring.

\begin{table*}[t]
\caption{Ablation Study of SpO$_2$ Estimation Performance under Different Sampling Rates and Input Settings}
\begin{center}
\begin{tabular}{|c|c|c|c|c|c|c|c|}
\hline
\textbf{Sampling Rate} & \textbf{Motion.} & \textbf{Perfusion.}
& \textbf{MAE ± SD} & \textbf{RMSE ± SD} & \textbf{PCC}
& \textbf{RMSE $\geq$85\%}
& \textbf{RMSE $<$85\%} \\
\hline

\multirow{4}{*}{100 Hz}
&  &  & 2.464 ± 1.067 & 3.307 ± 1.712 & 0.666 & 3.168 & 7.298 \\
\cline{2-8}
& \checkmark &  & 2.405 ± 1.021 & 3.184 ± 1.595 & 0.686 & 3.057 & 7.019 \\
\cline{2-8}
&  & \checkmark & 2.401 ± 1.096 & 3.237 ± 1.723 & 0.667 & 3.123 & 6.628 \\
\cline{2-8}
& \checkmark & \checkmark & \textbf{2.327 ± 1.080} & \textbf{3.098 ± 1.647} & \textbf{0.687} & \textbf{2.988} & \textbf{6.514} \\
\hline

\multirow{4}{*}{25 Hz}
&  &  & 2.438 ± 1.042 & 3.319 ± 1.683 & 0.661 & 3.173 & 7.670 \\
\cline{2-8}
& \checkmark &  & 2.382 ± 1.049 & 3.219 ± 1.687 & 0.684 & 3.079 & 7.698 \\
\cline{2-8}
&  & \checkmark & 2.370 ± 1.082 & 3.228 ± 1.701 & 0.663 & 3.108 & 6.960 \\
\cline{2-8}
& \checkmark & \checkmark & \textbf{2.305 ± 1.113} & \textbf{3.117 ± 1.743} & \textbf{0.685} & \textbf{3.007} & \textbf{6.804} \\
\hline

\end{tabular}
\label{tab:spo2_sampling_input_ablation}
\end{center}
\end{table*}

\subsection{Performance under Different Motion Levels}

\begin{figure}[b]
    \centering
    \includegraphics[width=\linewidth]{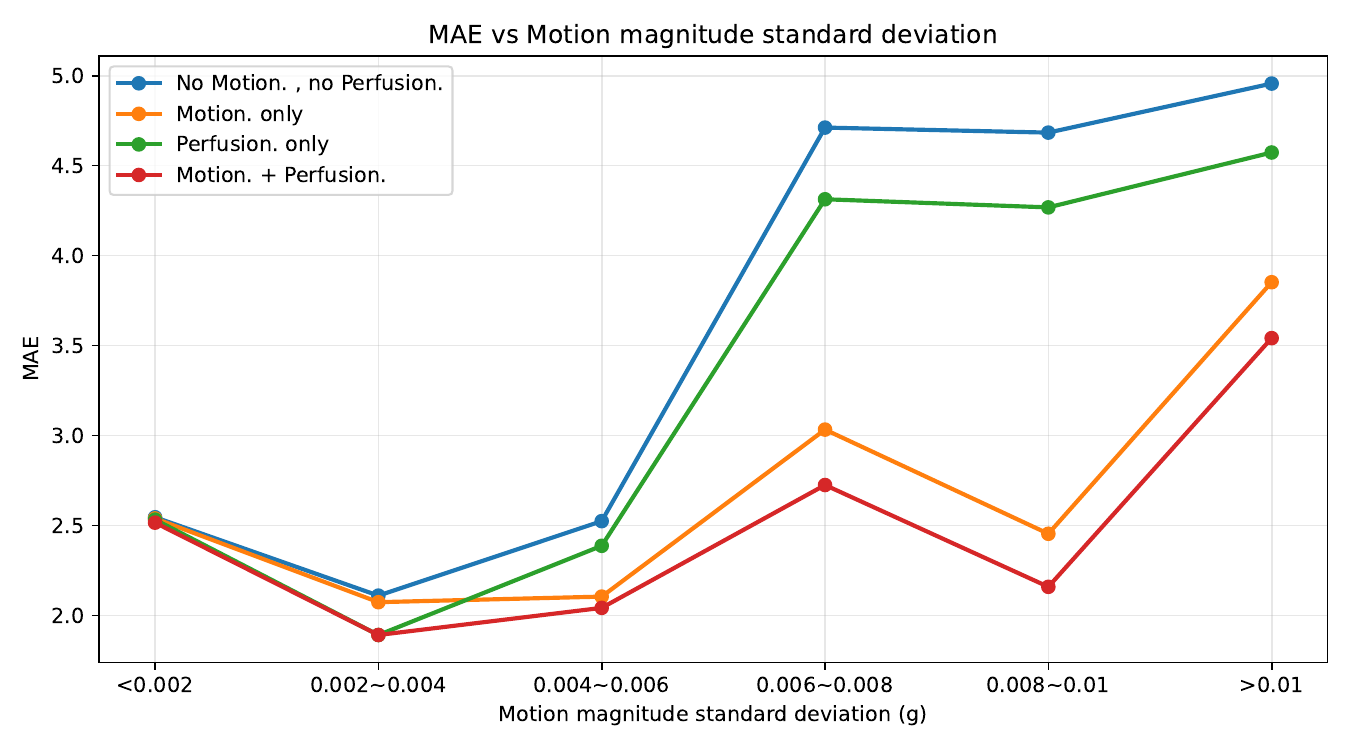}
    \caption{MAE under different motion levels.}
    \label{fig:motion_mae}
\end{figure}

To evaluate the effect of motion on SpO$_2$ estimation, we grouped the test windows according to the standard deviation of the accelerometer-derived motion magnitude and computed the MAE within each group. 
As shown in Fig.~\ref{fig:motion_mae}, the baseline method without motion or perfusion information shows increased error as the motion magnitude standard deviation becomes larger, indicating that wrist motion can degrade the reliability of wrist-based SpO$_2$ estimation.

Perfusion-guided calibration reduces the MAE across most motion groups compared with the baseline, suggesting that accounting for subject-specific perfusion status improves the stability of the R--SpO$_2$ mapping. 
In the higher-motion ranges, motion-aware beat selection further reduces the MAE by down-weighting beats that are more likely to be affected by motion-induced perturbations. 
Overall, combining motion-aware beat selection and perfusion-guided calibration achieves the lowest MAE across all motion levels.

\section{Conclusion and Future Work}

This work presented a lightweight framework for low-rate wrist-based SpO$_2$ estimation under micro-perturbations using motion-aware beat selection and perfusion-guided calibration. Instead of relying solely on a fixed calibration curve, the framework uses accelerometer-derived reliability weights to reduce the influence of micro-perturbations and incorporates a subject-specific reference perfusion index to improve calibration stability across different subjects.

Experimental results on a private wearable dataset showed that the proposed method achieved the best overall performance, with improvements from both motion-aware beat selection and perfusion-guided calibration. The 25-Hz setting achieved performance comparable to 100 Hz while reducing PPG sensor power consumption, supporting firmware-compatible and energy-efficient continuous wrist SpO$_2$ monitoring.

In future work, we will expand the dataset to include more participants and more diverse daily activity conditions. We will also explore dynamic adaptation of the reference perfusion index for long-term subject-specific monitoring.

\bibliographystyle{ieeetr}
\bibliography{EMBC}

\end{document}